\documentclass[runningheads]{llncs}
\usepackage{graphicx}
\usepackage{caption}
\usepackage{subcaption}
\usepackage[hyphens]{url}
\usepackage{hyperref}

\newcommand{\module}[1]{\paragraph{\normalfont \textbf{#1}}}

\usepackage[scaled=.75]{beramono}
\usepackage[T1]{fontenc}
\newcommand{\code}[1]{\texttt{#1}}
\newcommand{\of}[1]{_\mathrm{#1}}
\graphicspath{{./fig/}}

\begin{document}
\title{Multi-Scale Simulation Modeling for Prevention and Public Health Management of Diabetes in Pregnancy and Sequelae}
%
%
\author{        
Yang Qin\inst{1}
\and 
Louise Freebairn\inst{2,3,4}
\and
Jo-An Atkinson\inst{5,6,7}
\and Weicheng Qian\inst{1}
\and 
Anahita Safarishahrbijari\inst{1}
\and 
Nathaniel D Osgood\inst{1}
}
\authorrunning{Qin, Y. et al.}
%
\institute{ University of Saskatchewan \\
\email{first.last@usask.ca}
\and 
ACT Health, Canberra, Australia
\and
The Australian Prevention Partnership Centre, Sydney, Australia
\and
University of Notre Dame, Sydney, Australia \\
\email{Louise.Freebairn@act.gov.au}
\and
Decision Analytics, The Sax Institute, Australia 
\and
Sydney Medical School, University of Sydney, Australia
\and
Translational Health Research Institute, Western Sydney University, Australia \\
\email{jo-an.atkinson@saxinstitute.org.au}
}
\titlerunning{Simulation Modeling for Diabetes in Pregnancy and Sequelae}

\maketitle              
\begin{abstract}
Diabetes in pregnancy (DIP) is an increasing public health priority in the Australian Capital Territory, particularly due to its impact on risk for developing Type 2 diabetes.  While earlier diagnostic screening results in greater capacity for early detection and treatment, such benefits must be balanced with the greater demands this imposes on public health services. To address such planning challenges, a multi-scale hybrid simulation model of DIP was built to explore the interaction of risk factors and capture the dynamics underlying the development of DIP. The impact of interventions on health outcomes at the physiological, health service and population level is measured. Of particular central significance in the model is a compartmental model representing the underlying physiological regulation of glycemic status based on beta-cell dynamics and insulin resistance. The model also simulated the dynamics of continuous BMI evolution, glycemic status change during pregnancy and diabetes classification driven by the individual-level physiological model. We further modeled public health service pathways providing diagnosis and care for DIP to explore the optimization of resource use during service delivery. The model was extensively calibrated against empirical data.

\end{abstract}
\section{Introduction}
Gestational diabetes mellitus (GDM) is an increasing public health priority in the Australian Capital Territory (ACT), particularly on account of its impact on the risk of Type 2 Diabetes (T2DM) across the population \cite{Ferrara:2007aa,LEE2008124}. The increase of GDM is associated with increasing prevalence of risk factors including advanced maternal age \cite{Lao:2006aa}, obesity \cite{Athukorala_2010}, and sedentary behavior, growing GDM risk factors in those with family history of diabetes, and a growing number of residents whose ethnic background has traditionally been subject to elevated rates \cite{Ferrara:2007aa}.

Mathematical models characterizing diabetes progression, glucose hemostasis, pancreatic physiology and complications related to diabetes have been built by many researchers \cite{Ajmera_2013,TOPP2000605}. De Gaetano et al. \cite{De_Gaetano_2008} formulated a model representing the pancreatic islet compensation process, related to insulin resistance, beta-cell mass and glycemia (G) of a diabetic individual. Hardy et al. \cite{Hardy} proposed a model, characterizing mechanisms of anti-diabetic intervention and the corresponding impact on glucose homeostasis. Lehmann and Deutsch \cite{LEHMANN1992235} modeled the physiology underlying the interaction between insulin sensitivity (K$\of{xgI}$) and G of an individual with Type 1 diabetes (T1DM).  

Health simulation models commonly apply one of three types of modeling techniques: system dynamics modeling (SDM), agent based modeling (ABM) and discrete event simulation (DES). SDM captures and describes complex patterns of feedback and accumulation by solving sets of differential equations. While SDM can be applied at different scales \cite{Osgood_2007}, it is most commonly applied at the aggregated level, and its core components include the accumulation of elements (stocks), rate (flows), causal loops involving stocks (feedback), and delays \cite{Homer_2006,Kreuger}. By contrast, ABM simulates complex social dynamics by characterizing emergent system behavior as the result of within-environment interactions between individual elements in a system that are referred to as agents. ABM readily captures heterogeneous characteristics of agents, including agent history, situated decision making, structured interaction between agents typically evolving along multiple aspects of states and transitions and aggregation of individual outcomes \cite{Osgood_2007,Luke_2012}. DES characterizes individual-level, resource-limited progression through structured workflows which often associated with service delivery, queuing processes, waiting times and lists and resource utilization \cite{MARSHALL20155}.

Previous studies examining the health burden of GDM and its risk factors have predominantly relied upon cohort studies, administrative data or clinical trials \cite{XIONG2001221,Knight_Agarwal_2016,Feig229}. While filling a key set of research needs, given the dynamically complex nature of the interactions including feedback, accumulations, delays, heterogeneity, and interacting factors across many levels, it is difficult to use such studies to answer ``what-if'' question related with the risk factors and effects of interventions, particularly counter-factual whose outcomes have not yet been observed. Given the long time scales involved, cost, logistics, and ethical concerns, clinical trial studies may not be feasible for providing timely evaluation of novel portfolios of clinical-level and population-level interventions (PLI).  

In this work, we built a multi-scale hybrid model in AnyLogic (version 8.3.3) including SDM, ABM, and DES, to describe the dynamics of glycemic regulation (DGR), weight status and pregnancy, and to evaluate impacts of the interventions on DGR. While leaving most aspects of examination of model health findings to other forthcoming contributions, this paper introduces the design and structure of the model, provides illustrations of some of the types of interventions that the model can capture and simulation outputs.

The structure of the remainder of the paper is as follows: Model overview section describes the model structure and the simulation description. The next section briefly discusses model calibration and assumption. The model formulation section then describes the statecharts, DGR, weight dynamics, interventions, service delivery and offspring outcomes by hyperglycemia. Part 3 and 4 illustrate and discuss some of the model outputs and limitations of the model.

\section{Methods}

\module{Model Overview}
The \code{Person} class of the ABM includes the individual level characteristics such as evolving states, actions that change them, and the rules to trigger those actions (all captured in statecharts), parameters and functions.  
The ABM further represents family structure, weight at birth and evolution over the adult life course, individual history, inter-generational family context, pregnancy and diabetes classification, and implementation of the PLI. 
The SDM describing the DGR forms a sub-model encapsulated in the \code{Person} class. By encapsulating this SDM in the ABM structure, the model can capture side-by-side both individual characteristics and their evolution and continuous dynamics of the glucose-insulin system. The clinical service pathway for pregnant women in the ACT is described by a shared (global) DES, building on top of the ABM. The model will be discussed in its essentials in the following sections. Added elements of detail, the technical description associated with the model, are listed in the supplementary material, \url{https://www.cs.usask.ca/faculty/ndo885/GDM-ACT}, other material will be available later.

The model simulates a population of 200,000 female agents, each an instance of \code{Person} class. During the simulation, the agents can become pregnant, thereby experiencing the risk of GDM, and subsequently give birth, influencing the weight status and DGR of their descendants. The second generation agents also have their life-course shaped thoroughly by model dynamics. The information available for the descendants is, therefore, richer than in the initial population. Thus, the simulation requires a burn time of 60 years. 

\module{Model Calibration}
To estimate poorly- or non-measured parameters and to support the projection of status quo future incidence of DIP using model outputs, we calibrated a baseline model without interventions against the following historical data: the incidence of DIP of each ethnicity in ACT from 2008-2016, the prevalence of macrosomia by DIP status of in ACT from 2010-2016. To further capture the effects of inter-generational transfer of risk for GDM and T2DM which has been recognized from multi-generational epidemiological studies \cite{Franks460,Dabelea2208}, and occurrence of later-life diabetes, we drew on data related to developing diabetes by age 30 of offspring by their mother's G status (e.g., GDM, T2DM) and birth weight from population-wide administrative data from Saskatchewan, Canada. Birth weight can serve as an important marker of control of G in utero, and epidemiological studies suggest that it further influences the tendency towards GDM. The details of model calibration, e.g., objective function, will be available in other contributions.


\module{Model Formulation}
We discuss here several aspects of model formulation, particularly concentrating on statecharts, which encapsulate a discrete set of collectively exhaustive and mutually exclusive (lowest-level) states with respect to particular concerns, the actions by which the individual transitions between such states, and the rules under which such actions take place. 

\begin{figure}
\centering

\begin{minipage}{.6\columnwidth}
  \centering
  \includegraphics[scale=0.36]{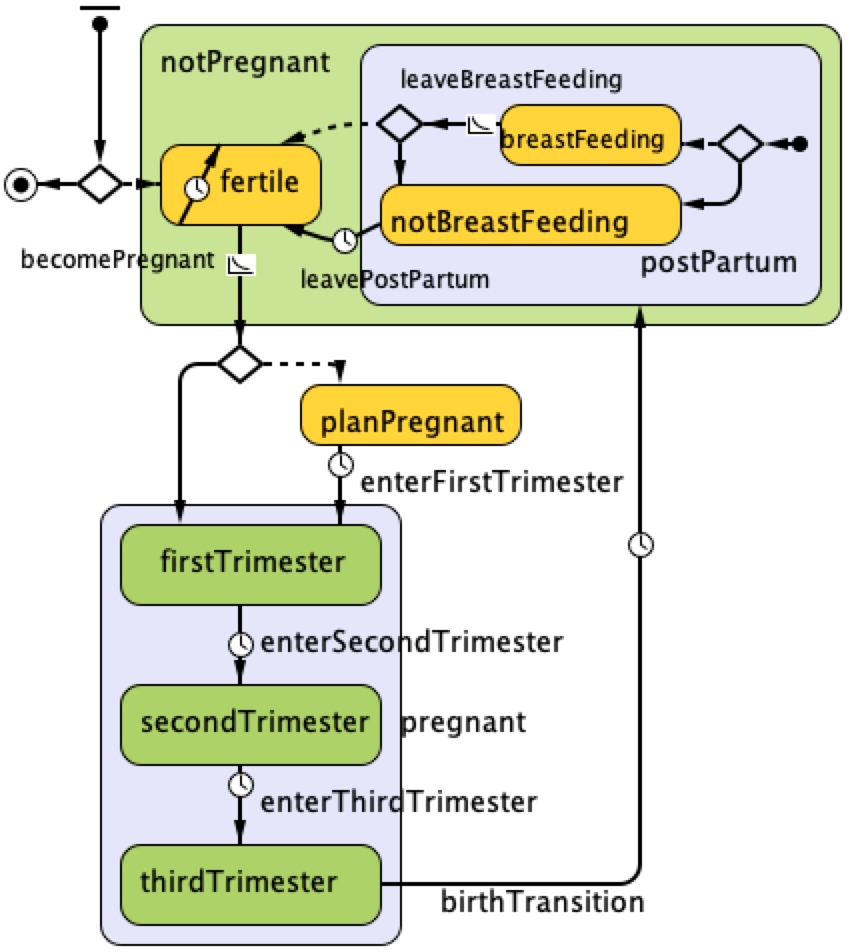}
  \captionof{figure}{Pregnancy statechart}
  \label{fig1}
\end{minipage}%
\begin{minipage}{.4\columnwidth}
  \centering
  \includegraphics[scale=0.36]{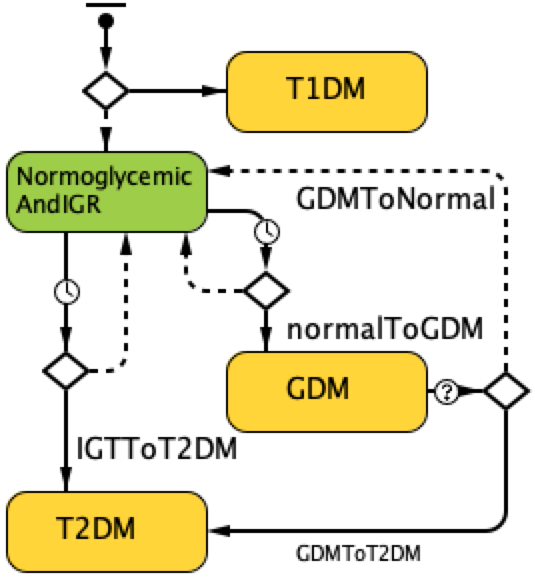}
  \captionof{figure}{Dysglycemia classification statechart}
  \label{fig2}
\end{minipage}

  

\end{figure}

Pregnancy statechart (Figure \ref{fig1}) indicates whether an agent is pregnant or not, and their transitions through different stages of pregnancy. Female agents with ages between 15 and 50 can transit between the \code{notPregnat}, \code{planPregnant}, and \code{pregnant} states. Agents in the \code{pregnant} hierarchical state will be in one of three substates, corresponding to trimesters of pregnancy. \code{notPregnant} agents will either be in the \code{fertile} or \code{PostPartum} state. \code{PostPartum} agent will either be in the \code{breastFeeding} or \code{notBreastFeeding} state. Of these, two of seven state transitions are memoryless transitions driven by a hazard rate (henceforth known as rate transitions), \code{becomePregnant} and \code{leaveBreastFeeding}; the hazard rate for \code{becomePregnant} is an age-ethnicity-specific fertility rate \cite{fertilityRate}. While the others are timeout transitions triggered after a specified residence time. The timeout transition \code{birthTransition} is particularly notable, as it introduces a new agent into the model.

Population statechart separates the population into three categories, initial female population, female descendant and male descendant.
Agents are initialized with different ages \cite{populationpyramid}, and assigned their ethnicity according to ACT demographic information taken from the 2011 Australian Census and National Health Survey. 
Type of ethnicity includes Australian Born, Australasian Diabetes in Pregnancy Society at risk group (ADIPS) \cite{Nankervis_2014}, Aboriginal and Torres Strait Islander (ATSI) and Other. All male descendants are excluded during simulation, and female agents leave the model upon reaching age 50. The female agents aged less than 50 leave the model by the age-specific death rate \cite{deathRate}.

Dysglycemia classification statechart (Figure \ref{fig2}) divides the G of an agent into four categories: \code{T1DM}, \code{NormoglycemicAndIGR}, \code{T2DM} and \code{GDM} states, according to clinical classification categories. Agents can occupy one of four states, and switch states by checking whether the G of agents exceed the threshold of each state (known as condition transition). Reflecting the fact that residence in the \code{GDM} state is only an option during pregnancy, pregnancy status is also considered. 
The \code{GDM} agents will either be in \code{NormoglycemicAndIGR} or \code{T2DM} state after pregnancy. Thresholds for \code{T2Dm} and \code{GDM} state are denoted as $G_{T2DM}$ and $G_{gdm}$, which are calculated by $C_{T2DM} \times G_{t}$ and $G_{t} \times C_{T2DM} \times C_{gdm}$, respectively, where $C_{T2DM}$, $C_{gdm}$ and $G_{t}$ are calibrated and equal to 1.636, 0.642 and 5.504, respectively.

DGR, the interaction between beta-cells, G and K$\of{xgI}$, is represented as an SDM based on the ordinary differential equation models of diabetes progression by De Gaetano et al. \cite{DeGaetano_2014,De_Gaetano_2008} and Hardy et al. \cite{Hardy}. To improve model scalability, a cyclic timeout event with time interval (\code{dt}) is used to solve the compartmental equations in SDM \cite{DeGaetano_2014,De_Gaetano_2008,Hardy}. 
Another cyclic timeout event with a time interval of 5 and 30 days during pregnant and non-pregnant periods, respectively, updates G and K$\of{xgI}$ using the Newton-Raphson method and other components in SDM. Parameter and function details are listed in the supplementary material.

To capture dynamics of K$\of{xgI}$ in different trimesters of pregnancy, postpartum and different weight status, respectively, we modified the (exogenous) equations giving K$\of{xgI}$ over time introduced by De Gaetano et al. \cite{DeGaetano_2014,De_Gaetano_2008},  
The model assumes the diminished K$\of{xgI}$ in pregnancy will gradually recover during the postpartum period to the value it would have held absent the pregnancy.
The model further assumes the K$\of{xgI}$ of overweight and obese agents would decline faster over age than that of agents with normal weight. 
In addition to the modification of equations giving K$\of{xgI}$, we modified the equation giving the spontaneous recovery rate of pancreas (\code{T$_{\eta}$}) for ADIPS. While ADIPS represents agents from a recognized high risk group with respect to GDM, the empirical data revealed that the ADIPS group actually had a higher proportion of healthy weight agents than that were present in the other groups, indicating that weight as a risk factor did not fully account for the higher risk levels.  Therefore, to capture the high incidence of GDM of ADIPS, the model assumes the \code{T$_{\eta}$} of ADIPS declines faster than that of other ethnic groups. Furthermore, to investigate effects on various types of intervention on the DGR, we incorporated the mechanism introduced by Hardy et al. \cite{Hardy} for the impact of lifestyle change (LC), metformin treatment (MT) and insulin treatment (IT) on K$\of{xgI}$. Elements of interventions making use of the LC, IT and MT are discussed in the next sections.


Weight dynamics are characterized as a continuous variable of BMI value, and a variable of Z-score of a BMI distribution (BMID), representing the position of BMI within the age group (AG) specific BMID. Upon entry to adulthood, agents are assigned a BMI value based on an AG specific BMID introduced by Hayes et al. \cite{Hayes_2015}, and its corresponding Z-score calculated by the BMI and mean of the BMID.  Hayes et al. \cite{Hayes_2015} reported that the BMID of the population within AG move toward higher BMI value through their life course. Applying an identical Z-score into the BMID of different AGs may position the agents into different weight categories. 
Therefore, for simplicity, the Z-score of agents are assumed to stay the same as they age, unless intervention or pregnancy \cite{Knight_Agarwal_2016} changes their BMI value and assigns a new Z-score to them. 
When an agent transfers from one AG to another, the BMI value of next AG of the agent will be calculated by applying the Z-score to the BMID of next AG in an event with a cyclic timeout of 10 years. As a continuous variable, another cyclic timeout event with an interval of 1 year is used to make the BMI of current AG change towards the BMI of next AG gradually. With this BMI-Z-score mechanism, we also captured BMI change following pregnancy; due to space considerations, the interested reader is referred to the supplemental material. Time-Varying weight distribution is required in light of the simulation burn time. We, therefore, employed importance sampling using an alternative BMID of female adults aged 25 to 64 years in 1980 and 2000 \cite{walls_2010} and AG specific BMID in 1995 and 2008 \cite{Hayes_2015}, to estimate the AG specific BMID in 1980. 

\begin{figure}
    \centering
    \includegraphics[scale=0.27]{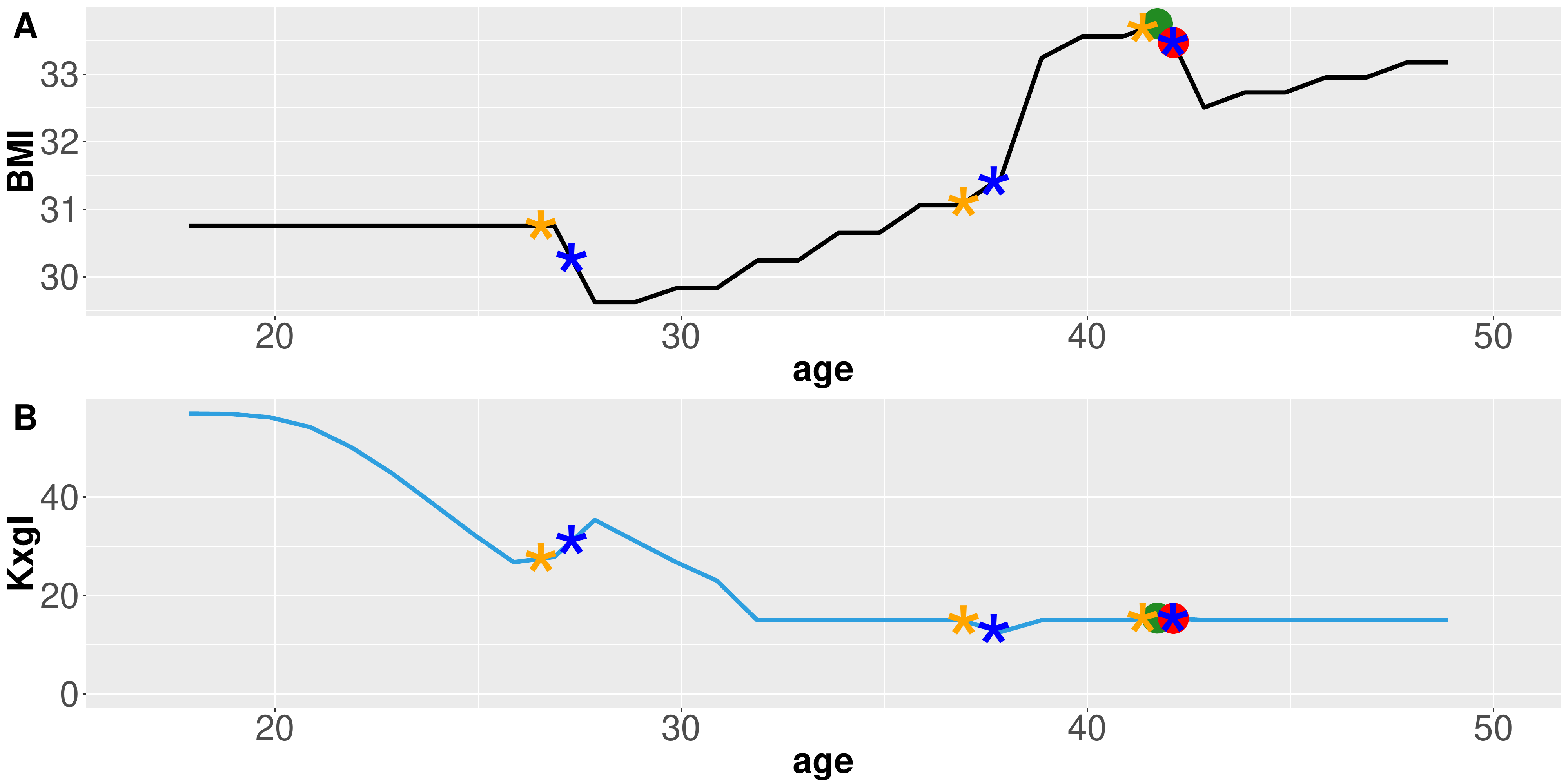}
    \caption{Illustration of the individual trajectory of BMI change (A) and K$\of{xgI}$ change (B) over age without PLIs and \code{Services}. Green and red dots are the start and end of the GDM period, respectively. Orange and blue stars are the beginning and the end of the pregnancy, respectively.}
    \label{fig7}
\end{figure}

\begin{figure}
    \centering  
    \includegraphics[scale=0.27]{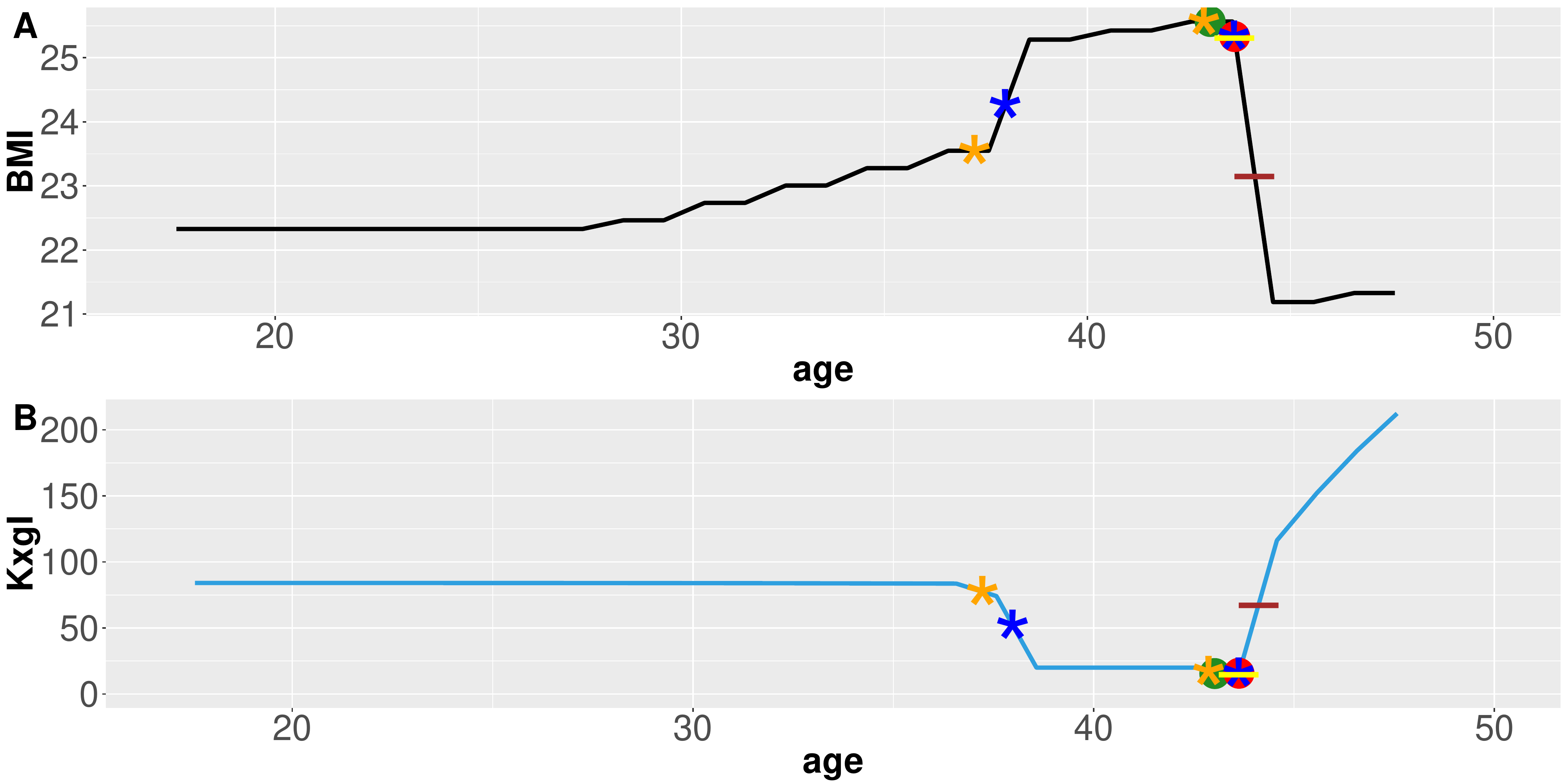}
    \caption{Illustration of the individual trajectory of BMI change (A) and K$\of{xgI}$ change (B) over age with DRI. Color Labels of GDM and pregnancy are same with Figure \ref{fig7}. Yellow and brown bars are the start and the end of the DRI, respectively.}
    \label{fig8}
\end{figure}


ACT clinical service pathway, \code{Services}, is modeled using DES. And in \code{Person}, a statechart reflects type of health care that an agent is currently being delivered, which is separated as the \code{InPrimaryCare} state reflecting that a non-pregnant woman is receiving usual health care services through a general practitioner, and the \code{InACTHealthService} state reflecting that a pregnant woman is moving through the \code{Services}. The DES and statechart not only models the effects of LC and IT in reducing the risk of progression to T2DM after delivery and implemented PLIs in the \code{InPrimaryCare}, but also leave room for investigating resource use and costs associated with service provision. The blocks (i.e. \code{dipAssessment}, \code{antenatalCare}, \code{dieticianReview} and \code{lifestyleOrInsulinTreatment}) in the \code{Services} form a sequence of operations, providing pregnant agents the DIP assessment test, education of LC, IT for the agents with DIP, and further deliver postpartum checks to agents. Details of the DES are described in the supplemental material.

PLI includes consideration of a public health messaging and mobile app support intervention (PHMMASI), health professional support intervention (HPSI), diet review intervention (DRI), and public health messaging and support intervention (PHMSI). The difference between the \code{Services} and the PLI is that PLIs are initiated during the non-pregnant period, while the \code{Services} is triggered during pregnancy. All PLIs share a similar mechanism of taking optional LC and BMI reducing. Specifically, overweight and obese agents reduce their BMI, drawing the extent of that reduction from a normal distribution, while the normal weight agents keep their BMI invariant.  The interventions are variants of each other with respect to who they are target, the intervention triggering time, and the length and strength of adherence of the LC. In detail, the agents with age between 20-35 take the PHMMASI and retake it according to certain probability \cite{Flores_Mateo_2015}.
The HPSI takes place at the \code{planPregnant} state, and works on women with risk factors, e.g., BMI $>$ 28, age $>$ 30, ADIPS ethnicity \cite{WEISMAN2011265}. 
DRI and PHMSI both take place between pregnancies, and target on women who had DIP in previous pregnancies and on women who have given birth, respectively. For the HPSI and PHMSI, the adherence and length of LC are flexible, whereas the agents who are subject to DRI take mandatory, lifelong, strongly adherent LC.

The outcomes for baby and mother including birth weight (e.g., macrosomia), type of birth (e.g., Caesarean section), NICU admission, and shoulder injury, are triggered in \code{birthTransition} of the pregnancy statechart. The probability of occurrence of baby outcomes was calculated based on the study by The HAPO Study Cooperative Research Group \cite{HAPO_2008}. Furthermore, information of the mother is passed on to the new child, including DIP status, age, weight status and ethnicity. The mother's DIP status influences the K$\of{xgI}$ of the child by multiplying a coefficient to the K$\of{xgI}$ calculated by the modified equations giving K$\of{xgI}$ over time.

\section{Results}
We show here several scenarios that demonstrate the functioning of the model at the level of an individual's health history. Figure \ref{fig7} A and B show the individual trajectories of BMI changes and K$\of{xgI}$ over age without the PLIs and the \code{Services}, respectively. Figure \ref{fig7} A illustrates the agent entered adulthood with a BMI of 30.75, and her BMI reduced over one BMI unit after the first and third pregnancy and three BMI units after the second pregnancy. Other than pregnancy, Figure \ref{fig7} A also demonstrates continuous BMI change over age. The K$\of{xgI}$ remained constant under the age of 18 but declined over time according to the value of BMI after the age of 18. The agent developed GDM at the third pregnancy, as shown by the dots in Figure \ref{fig7}. Furthermore, Figure \ref{fig7} B reflects the decreasing K$\of{xgI}$ during pregnancy and recovery in postpartum. We can see from Figure \ref{fig7} that K$\of{xgI}$ is declining in parallel to, and, in fact, in response to the increase in BMI.

From Figure \ref{fig8}, we can see that the agent increased over one BMI unit after the first pregnancy, and K$\of{xgI}$ was decreased in response to this BMI increase. At the second pregnancy, the agent developed GDM but retained their BMI and corresponding Z-score after pregnancy. But the DRI reduced that agent's BMI and significantly increased  K$\of{xgI}$ from 20 to 116 at the end of BMI reduction period (6 months), following which the K$\of{xgI}$ continued to increase due to strong adherence in LC, as shown by the bar labels in Figure \ref{fig8} A and B.

\section{Discussion}
This paper has described a novel multi-scale model that utilizes three types of system simulation methods to provide a versatile, powerful and general platform for examining interventions to address the growing epidemic of GDM and T2DM in the ACT. The model achieves such versatility by virtue of maintaining a core underlying physiological representation that captures the common generative pathways mediating diverse needs in the model, to capture effects of lifestyle and clinical interventions, to capture clinical categorization, to represent the effects of each of pregnancy, aging and BMI change, and the longer term-effects of one pregnancy (via beta-cell mass and function) on later pregnancies and subsequent material risk of T2DM, and outcomes of interest. Such a representation can also flexibly capture the impacts of maternal status on the offspring.  

A high level of heterogeneity at the individual level, e.g., family context, risk factors for diabetes and life course trajectories motivated the use of ABM as the core component of this hybrid model. The ABM permits a high-resolution representation of relevant dynamics of individual objects and further allows the implementation of finely targeted interventions. Compared to ABM, SDM simulates a system in a more abstract and general way. The high level of abstraction of DGR makes it a suitable candidate for SDM. The \code{Services} can be described as a sequence of operations, DES, therefore, was selected to model the \code{Services}, and to study the resource allocation and effect of clinical interventions.

While empirical models of necessity represent simplifications of processes in the world, the model here includes a requisite degree of detail to capture a remarkably broad set of factors. Nonetheless, they remain important limitations in the model that are ripe for addressing. These notably include a lack of detail with regards to childhood dynamics (including weight change), neglect to social network effects on behavior, and an overly simple representation of changes in K$\of{xgI}$ and behavior change. Extensions of the model to capture such effects, and to capture cost and resource components of scenarios, remain an important priority. 
%
%
%
\bibliographystyle{splncs04}
\bibliography{sbp2019_gdm}

\end{document}